# Binary Weighted Memristive Analog Deep Neural Network for Near-Sensor Edge Processing

O. Krestinskaya[1], and A. P. James[1]
[1]Nazarbayev University, Astana, Kazakhstan, email: apj@ieee.org

**Abstract**—The memristive crossbar aims to implement analog weighted neural network, however, the realistic implementation of such crossbar arrays is not possible due to limited switching states of memristive devices. In this work, we propose the design of an analog deep neural network with binary weight update through backpropagation algorithm using binary state memristive devices. We show that such networks can be successfully used for image processing task and has the advantage of lower power consumption and small on-chip area in comparison with digital counterparts. The proposed network was benchmarked for MNIST handwritten digits recognition achieving an accuracy of approximately 90%.

## I. Introduction

The use of the resistive switching memories in crossbar provides an option to build analog computing neural networks to implements dot product operation [1]. While, there are dot product implementation using the memristive crossbar for analog neural network [2], the realistic implementation of such system is difficult to achieve due to the limitation of the number of resistance levels in the memristors. Further, in most of the previous studies only the conceptual implementation of the forward propagation with ideal components is reported [2]. There is a lack of the studies showing the implementation of the neural network considering non-idealities of real hardware components and the effects of real memristors on the performance of the neural network has not been investigated yet. Therefore, the analog hardware implementation of the neural networks is still an open problem.

Binarized Neural Network (BNN) [2] can be one of the possible practical alternative solutions to the problem of analog hardware implementation of the neural network. BNN is the neural network with binary weights, binary activation function, binary inputs or combined approached. Recently, several variations of the software implementation of BNN algorithms have been reported, while the analog hardware implementation of BNN system remains an open problem. The implementation of such network can be useful for increasing the speed and enhancing the performance of near-sensor visual data classification in edge devices. In addition, due to the limitations of the levels of the memristive devices, BNN hardware implementation can be built with the existing binary memristive devices. The memristive devices have been proven to be efficient for various architectures in terms of scalability, power consumption and on-chip area [3,4,5,6,7,8,9,10,11,12].

In this paper, we propose the analog hardware implementation of BNN with backpropagation and binary weights that can be scaled to implements deep BNN. We show the circuit level implementation of analog circuits involved in the BNN architecture. The overall performance of the system is tested for different number of neural network layers. The increase in the number of hidden layers and connections between them represented by the memristive crossbar and analog implementation of different activation functions allows to build binary deep neural network. The proposed network is tested using MNIST and IRIS databases.

## II. Hardware implementation of BNN

There are several implementations of BNN with digital logic [2,3]. In this work, we focus on the analog implementation of BNN with binary weights trained with the backpropagation algorithm. In the training stage, the weights are updated with the backpropagation algorithm with gradient descent. After the training, the weights are binarized before the classification process. Such binarization is useful for the hardware implementation of the BNN using memristive crossbar approach, where only high and low value of memristor is possible to achieve with the current technology.

In this paper, the backpropagation algorithm is similar to the backpropagation with gradient descent shown in [4]. However, after the calculation of the updated weight $w$, it is binarized to either high $w_{high}$ and low values $w_{low}$, or their negative alternatives $-w_{high}$ and $-w_{low}$. All weights in each weight matrix are rounded to the closest positive or negative weight. Such binarization is useful for the hardware implementation of the BNN using memristive crossbar approach, where only high and low values of memristor are possible to achieve and for the memristive devices, where only Ron and Roff values are stable.

The proposed hardware implementation of the BNN is shown in Fig.1. The design consists of the memristive crossbars used as storage of weights, activation function, sequence control unit, storage and normalization unit, weight sign control unit, neural network training unit and weight update unit. Each input image is reshaped and feed into the crossbar rows to Vin1 and up to Vink, where k is the number of pixels in a single image. The number of the rows in the crossbar corresponds to the number of the inputs neurons to the crossbar and the number of columns refers to the number of output neurons in the BNN. The output of the crossbar is the current in the transistors connected to the columns, which represents dot product operation. The currents from each column of the crossbar are read sequentially one column at a time.

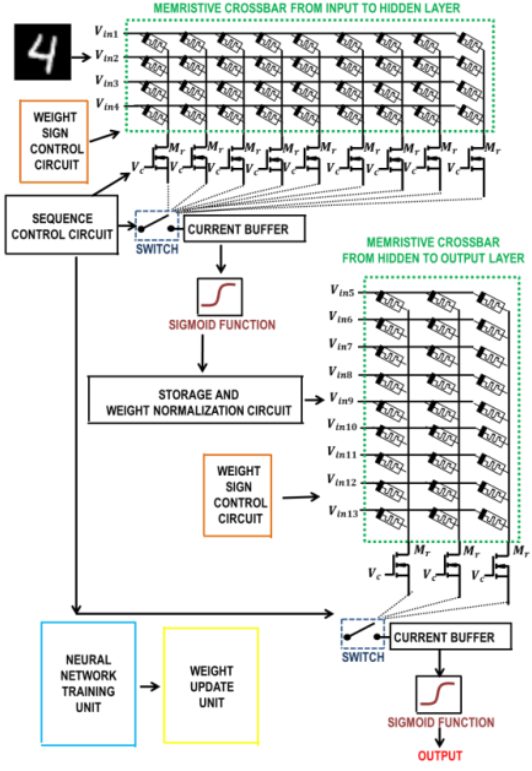

Fig.1. Overall architecture of the proposed BNN implementation

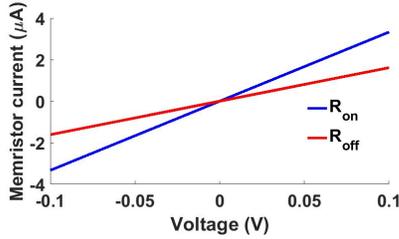

Fig.2. Output current range for a single memristor for Ron and Roff.

The design of BNN is based on the binary memristive crossbar representing the weights between the layers of the neurons in the neural network. The output of the crossbar is the current in the transistor $M_r$. The currents from each column of the crossbar are read sequentially one column at a time. The transistors are controlled by the control signal $V_c$, which switches the transistors ON and OFF. To ensure that the power consumption of the circuit is not too high, the maximum $V_c$ switches on the transistor is restricted to 1V. In order to keep the current through the transistor proportional to the applied voltage signal, the following condition should be fulfilled: $V_{DC} \square V_{GS} - V_t$, where $V_t$ is the transistor threshold. We adjust the overall circuit to TSMC 180nm CMOS process, and the NMOS transistor that is used in the crossbar has the threshold level $V_t = 0.35V$. Therefore, the maximum drain voltage of the transistor to ensure linear operation is $V_D = 0.65V$ and $V_{in} < 0.65V$. For this work, the HP memristor model is used and the analog circuits and transistor parameters are adjusted to the following values of memristors: Ron = 3k$\Omega$ and Roff = 62k$\Omega$ with the threshold level of 1V. To avoid the summation of the current to exceed the range of the sigmoid function, the input voltage levels are normalized between 0V and 0.1V. This range was obtained by analyzing the range of output currents flowing through a single memristor for Ron and Roff regions shown in Fig.2.

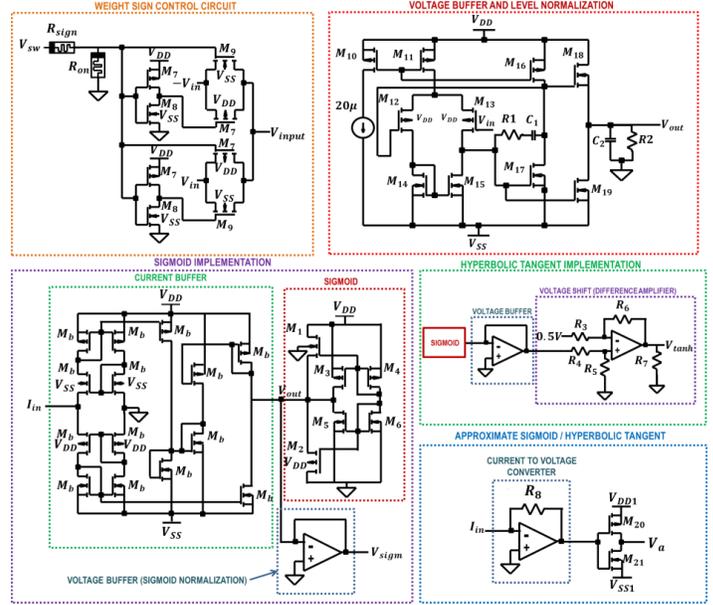

Fig.3. Analog circuit components used in BNN, including weight sign control circuit, voltage buffer and level normalization, sigmoid circuit implementation and additional activation functions, such as approximate tangent and sigmoid.

Fig.3 shows the analog circuit implementation of the BNN. As in the binarization algorithm the neural network weights can be both positive and negative, and it is not possible to have a negative value for the resistance of HP memristor, the weight sign control circuit is implemented. The weight is stored in the memristor $R_{sign}$. The output current from the crossbar is applied to the current buffer to read it without the effect of the following circuits on the crossbar performance. Next, the output of the current buffer is applied to the activation function circuit. The main activation function circuit for the BNN is a sigmoid. As we design the system for TSMC 180nm CMOS process with Vdd = 1.8V, the output of the sigmoid requires normalization, which is performed by the voltage buffer circuit. To implement analog deep BNN, the implementation of other activation functions is required, such as hyperbolic tangent, approximate sigmoid and approximate hyperbolic tangent shown in Fig.3. The design of the components allows using the same circuit to implement different functions with a simple switch. This can enhance the performance of the BNN without the implementation of the additional components.

## III. SIMULATION RESULTS

The system level simulations were performed in MATLAB for MNIST and IRIS database. Table I shows the performance evaluation of the proposed BNN, comparing to the conventional neural network (NN) with backpropagation. In general, the average accuracy results for BNN (3 layers) are lower than the conventional NN. However, the selection of a particular value of the weights allows to achieve the maximum possible accuracy, which is higher than the performance accuracy of the conventional NN. The overall performance of deep BNN (6 layers) and deep NN is lower, in comparison to the 3 layer NN and BNN. It can be improved by adjusting the simulation parameters. In this simulation, showed the worst case scenario, where the number of output and hidden layer neurons is limited. The main idea of the simulations is to show that the accuracy that can be achieved by BNN is similar to the conventional neural network accuracy with analog weights. However, the analog implementation of BNN with the real devices is easier to implement.

TABLE I. PERFORMANCE EVALUATION OF THE PROPOSED BNN COMPARING TO THE CONVENTIONAL NEURAL NETWORK (NN).

| Method | Average accuracy | | Maximum accuracy | |
|---|---|---|---|---|
| | MNIST | IRIS | MNIST | IRIS |
| Conventional NN | 89.9% | 96.6% | 91% | 100% |
| **Proposed BNN** | **89.6%** | **68%** | **100%** | **100%** |
| Conventional deep NN | 32% | 32% | 62% | 62% |
| **Proposed deep BNN** | **22%** | **10%** | **39.9%** | **32%** |

The circuit level simulations were performed in SPICE for TSMC 180nm CMOS technology. Fig. 4 shows the simulation results for the proposed circuits for the activation function implementation for the variation of the crossbar output current from −90μA to 90μA. Fig.4(a) shows the output Vout of the sigmoid circuit. Fig. 4(b) and Fig. 4(c) shows the output of the voltage buffer Vsigm that refers to the normalization of the sigmoid to 1V and 0.1V, respectively. Fig.4(d) shows the tanh function implementation refering to Vtanh. Fig.4(e) and Fig.8(f) illustrate the implementation of the approximate sigmoid and approximate tanh, respectively.

Fig. 5 shows the timing diagrams for the proposed circuits. Fig. 9(a) shows the input current to the current buffer from the transistor in the crossbar. Fig. 5(b) corresponds to the output current from the buffer, which is the input to the sigmoid circuit. Fig. 5(c), Fig.5 (d) and Fig. 5(e) illustrates the output of the sigmoid Vout, sigmoid normalization to 1V and the normalization to the 0.1V , respectively. Fig. 5(f) represents the output of the circuit for the tanh function implementation Vtanh . Fig.5 (g) and Fig.5 (h) refers to the implementation of rough sigmoid and tanh functions, respectively. In Fig.5 (d), Fig.5 (e) and Fig.5 (f), the small delay is caused by the capacitor. To enhance the quality of these outputs, the capacitor and resistor values in the OpAmp have to be adjusted.

The calculation of the power dissipation and on-chip area are shown in Table II. In addition, the total power dissipation and area calculations for the basic 3 layer BNN with sigmoid activation function and 2 crossbars of the size of 4 × 10 are shown. The power consumption and on-chip area of the deep BNN depends on the activation functions used in the network and the size of the hidden layers. The calculated total area and power are 4839.9 μm$^2$ and 1072.4mW, respectively. These results can be improved by introducing low power amplifier and using FinFET devices.

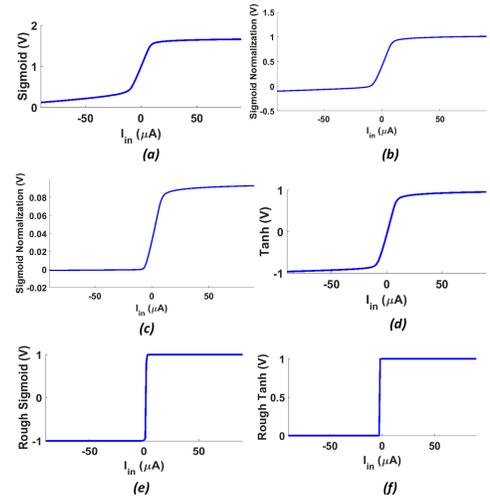

Fig. 4: Timing diagrams for the proposed circuits: (a) input current, (b) buffer output current, (c) sigmoid, (d) normalization of the sigmoid to 1V , (e) normalization of the sigmoid to 0.1V, (f) tanh function, (g) approximate sigmoid and (h) approximate tanh.

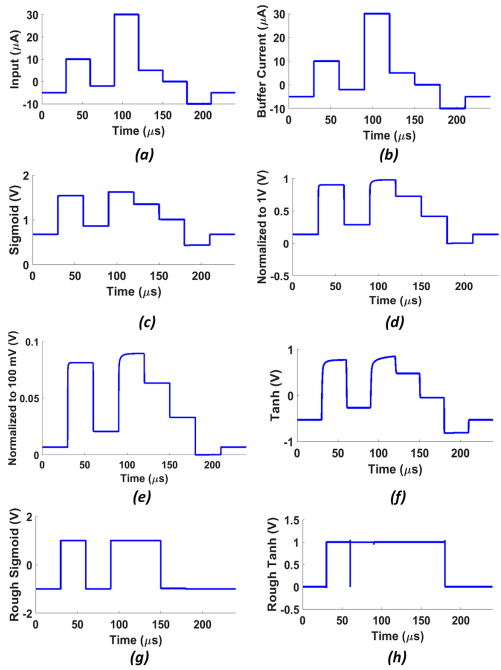

Fig. 5: Implementation of the activation functions for the variation of the crossbar output current: (a) sigmoid circuit output, (b) normalization of the sigmoid to 1V, (c) normalization of the sigmoid to 0.1V, (d) tanh function, (e) approximate sigmoid and (f) approximate tanh function.

TABLE II. POWER CONSUMPTION AND ON-CHIP AREA CALCULATION FOR THE CIRCUIT COMPONENTS.

| Circuit component | Power consumption | On-chip area |
|---|---|---|
| Crossbar (4 input neurons and 10 output neurons) | 5μW | 1.36μm$^2$ |
| Weight control circuit | 11.4μW | 7.98μm$^2$ |
| Sigmoid | 11.4μW | 184μm$^2$ |
| Current buffer | 149μW | 280μm$^2$ |
| Voltage buffer | 451μW | 1954.6μm$^2$ |
| Voltage shift (difference amplifier) | 3.952mW | 2581.4μm$^2$ |
| Approximate sigmoid/tanh | 41.19mW | 2118μm$^2$ |
| **Total for 3 layer BNN with sigmoid:** | **1072.4 mW** | **4839.9μm$^2$** |

## IV. DISCUSSION

As all circuits in the proposed BNN are analog, it can be useful for near sensor processing and can be integrated directly into sensor without additional conversion stages. As the ADC converters and sampling circuits are the main limitation of the processing speed in data processing sensors, the proposed fully analog architecture is one of the solutions to increase processing speed, which is important for large scale problems.

However, there are several drawbacks and open problems to be considered. As the current memristor technology is unstable and memristors can have switching problems, the selection of memristive devices for this architecture is important, and the effect of the non-ideal behavior of real memristive devices should be studied. The current memristive technology allows the implementation of binary states [13], however the endurance issues, effects of switching probability of the accuracy [14] and other aspect of non-ideal behaviour of memristive devices should be studied.

In addition, the design of the CMOS components in the proposed circuit can be optimized and improved to ensure scalability of the circuits, lower power consumption and smaller on-chip area. To ensure scalability of the circuits, the OpAmp circuit can be replaced with low power amplifier. To implement full architecture, the design of mixed signal sequence control circuit and is required. Also, the system level simulations should be performed to verify the number of hidden layer neurons and memristive weights to ensure the high performance accuracy.

## V. CONCLUSION

In this work, we present the analog hardware implementation of BNN that can be integrated for near-sensor data classification. We presented the application of the memristive binary crossbar for BNN and the design of the hardware units and the analog circuits for activation functions. The performance of the proposed BNN was evaluated using MNIST and IRIS database. The average accuracy that can be achieved with BNN is approximately 90%. The hardware implementation is scalable. Thus, it is possible to implement deep BNN. The area and power requirements for 3 layer BNN with the crossbar of the size of 4 × 10 are 1072.4 mW and 4839.9μm$^2$, respectively. As an extension to the reported results, we plan to include an identification of the variation of performance accuracy due to the instability of memristive technology and scalability of the crossbar, adjustment of deep BNN structure and activation functions to achieve better performance and higher accuracy.